\documentclass[10pt,journal,twoside,final]{IEEEtran}

\newtheorem{theo}{Theorem}

\newtheorem{cor}{Corollary}

\IEEEoverridecommandlockouts

\makeatletter
\newcommand{\biggg}[1]{{\hbox{$\left#1\vbox to 20.5pt{}\right.\n@space$}}}
\newcommand{\Biggg}[1]{{\hbox{$\left#1\vbox to 23.5pt{}\right.\n@space$}}}
\newcommand{\bigggg}[1]{{\hbox{$\left#1\vbox to 26.5pt{}\right.\n@space$}}}
\newcommand{\Bigggg}[1]{{\hbox{$\left#1\vbox to 29.5pt{}\right.\n@space$}}}
\newcommand{\biggggg}[1]{{\hbox{$\left#1\vbox to 32.5pt{}\right.\n@space$}}}
\newcommand{\Biggggg}[1]{{\hbox{$\left#1\vbox to 35.5pt{}\right.\n@space$}}}
\newcommand{\bigggggg}[1]{{\hbox{$\left#1\vbox to 38.5pt{}\right.\n@space$}}}
\newcommand{\Bigggggg}[1]{{\hbox{$\left#1\vbox to 41.5pt{}\right.\n@space$}}}
\makeatother

\usepackage{bm,cite,algorithm,algorithmic,float,amsmath,amssymb}
\usepackage[dvips]{graphicx}
\usepackage{subfigure,amsfonts}
\usepackage{mdwmath}
\usepackage{mdwtab}
\usepackage{xcolor,color}
\usepackage{cool}
\usepackage{balance}
\usepackage{diagbox}
\floatname{algorithm}{Algorithm}

\makeatletter
\renewcommand\paragraph{\@startsection{paragraph}{4}{\z@}%
            {-2.5ex\@plus -1ex \@minus -.25ex}%
            {1.25ex \@plus .25ex}%
            {\normalfont\normalsize\itshape}}
\makeatother
\usepackage{epstopdf}

\usepackage[style=0]{mdframed}
\usepackage{showexpl}

\mdfdefinestyle{exampledefault}{%
rightline=true,innerleftmargin=10,innerrightmargin=10,
frametitlerule=true,frametitlerulecolor=green,
frametitlebackgroundcolor=yellow,
frametitlerulewidth=2pt}

\allowdisplaybreaks

\begin{document}


\title{A  New Rate Splitting Strategy for Uplink CR-NOMA Systems}

\author{
Hongwu~Liu, 
Zhiquan Bai,  
Hongjiang Lei,  
Gaofeng Pan,  
Kyeong Jin Kim,  
and Theodoros A. Tsiftsis

\thanks{H. Liu is with the School of Information Science and Electrical Engineering, Shandong Jiaotong University, Jinan 250357, China (e-mail: liuhongwu@sdjtu.edu.cn).}
\thanks{Z. Bai is with the School of Information Science and Engineering, Shandong University, Qingdao 266237, China (e-mail: zqbai@sdu.edu.cn).}
\thanks{H. Lei is with the School of Communication and
Information Engineering, Chongqing University of Posts and Telecommunications, Chongqing 400065, China (e-mail: leihj@cqupt.edu.cn).}
\thanks{G. Pan is with the School of Cyberspace Science and Technology, Beijing Institute of Technology, Beijing 100081, China (e-mail: gfpan@bit.edu.cn).}
\thanks{K. J. Kim is with Mitsubishi Electric Research Laboratories, Cambridge, MA 02139 USA (e-mail: kkim@merl.com).}
\thanks{T. A. Tsiftsis is with the School of Intelligent Systems Science and Engineering, Jinan University, Zhuhai 519070, China (e-mail: theo\_tsiftsis@jnu.edu.cn).}
}

\maketitle
\setcounter{page}{1}
\begin{abstract}
In uplink non-orthogonal multiple access (NOMA) channels, the existing cooperative successive interference cancellation (SIC) and power control (PC) schemes lack the capability of achieving the full capacity region, which restricts the outage performance of uplink NOMA users. For the uplink cognitive radio inspired NOMA system, we propose a new rate splitting (RS) strategy to maximize the achievable rate of the secondary user without deteriorating the primary user's outage performance. Based on the interference threshold and its own channel gain, the secondary user adaptively conducts RS, transmit power allocation and SIC, which utilizes the transmit power efficiently. The closed-form expression for the outage probability is derived for the secondary user. Numerical results show that the proposed RS scheme achieves the best outage performance for the secondary user among the existing cooperative SIC and PC schemes.
\end{abstract}


\section{Introduction}
In power-domain non-orthogonal multiple access (NOMA) systems, multiple users' signals  are  multiplexed on the same time/frequency resource block, which  results  in inter-user interference (IUI) at the receiver side. Due to the powerful capability of eliminating IUI and simple structure, successive interference cancellation (SIC) has been utilized to detect the multiplexing signals in NOMA systems \cite{NOMA_SIC}. Nevertheless, the decoding performance of a NOMA receiver  strongly depends on how the SIC decoding order and the associated power control (PC) are optimized \cite{Unveiling_SIC_PartI,SGF_NOMA_QoS,Hybrid_SIC_NOMA_New,
NOMA_Dynamic_SIC,NOMA_Advanced_SIC,NOMA_backoff}.

Under the quality of service (QoS) oriented criterion, the users with more stringent QoS requirements are decoded first, while the others are decoded in the remaining stages of the SIC \cite{Unveiling_SIC_PartI,SGF_NOMA_QoS,Hybrid_SIC_NOMA_New}. Under the channel state information (CSI)-based criterion, the users with better CSI are firstly decoded before decoding the users with the worse CSI \cite{NOMA_SIC,NOMA_Dynamic_SIC,NOMA_Advanced_SIC,NOMA_backoff}. When a fixed SIC decoding order is applied in the uplink NOMA systems, the users decoded in the primary stages of the SIC are inevitably affected by the IUI, which may result in severe outage probability error floor. Fortunately, the recently proposed QoS-guaranteed SIC (QoS-SIC) scheme in \cite{Unveiling_SIC_PartI} and \cite{SGF_NOMA_QoS} avoids the outage probability error floor to some extent by adopting a hybrid decoding order, which takes into account both CSI and QoS requirements for calculating the users' priorities in SIC. Furthermore, a new hybrid SIC (NH-SIC) scheme was proposed in \cite{Hybrid_SIC_NOMA_New}, which introduced the PC method to increase the achievable rate.

However, the existing SIC and PC schemes cannot fully achieve the capacity region of the uplink multiple access channels (MACs) compared to the rate splitting (RS) \cite{Rate_Splitting_MAC,RS_NOMA_UL_fair}, hence only suboptimal outage performance can be achieved by the QoS-SIC and NH-SIC schemes. In this paper, we propose an RS strategy for the uplink cognitive radio (CR) inspired NOMA (CR-NOMA) system.
Without deteriorating the outage performance of the primary user compared to that in orthogonal multiple access (OMA), the secondary user splits its signal into two streams and allocates the transmit power efficiently to each stream to maximize the achievable rate. Benefiting from advantage of the RS for achieving the full capacity region of MACs, our proposed RS scheme attains the best outage performance for the secondary user.
To clarify the advantage of the proposed RS scheme, we carry out a comprehensive analysis and provide detailed comparisons over the existing SIC and PC schemes \cite{SGF_NOMA_QoS,Hybrid_SIC_NOMA_New}.

\section{System Model and RS Strategy}

\subsection{System Model}

We consider an uplink CR-NOMA system consisting of a primary user $U_0$, a secondary user $U_1$, and a base-station (BS). Each node is equipped with a single antenna. The channel coefficients from $U_0$ and $U_1$ to the BS are denoted by $h_0$ and $h_1$, respectively, which are modeled as independent and identically distributed (i.i.d.) circular symmetric complex Gaussian random variables with zero mean and unit variance. We assume that the channels follow a statistical fading. In addition, we
assume the same CSI acquisition procedure as that in \cite{Hybrid_SIC_NOMA_New} and take perfect SIC for comparing the system performance of the proposed RS scheme with the benchmarks \cite{SGF_NOMA_QoS,Hybrid_SIC_NOMA_New}.

Required by the stringent QoS requirements,  $U_0$ transmits at a fixed data rate. $U_1$ is allowed to share the same resource block with $U_0$ only when the $U_1$'s transmission does not deteriorate the $U_0$'s outage performance compared to the counterpart of $U_0$ in OMA. In other words, $U_0$ experiences the same outage performance as in OMA for delay-limited transmissions.

To begin with, we introduce $P_i$, $g_i$, and $\varepsilon_i$ for user $U_i$ ($i=0, 1$) to represent its transmit power, channel gain, and target rate related metric, respectively, where $g_i = |h_i|^2$ and $\varepsilon_i~=~2^{\hat R_i}-1$ with $\hat R_i$ denoting the target rate of $U_i$.
Prior to each transmission block, the BS sends an interference threshold to $U_1$ to guarantee the QoS requirements of $U_0$. Considering that $U_0$ can achieve the same system performance as in OMA, the interference threshold is given by
\begin{eqnarray}
    \tau = \max\left\{0, \frac{P_0 g_0}{\varepsilon_0}  -1  \right\}.
\end{eqnarray}
With respect to $\tau$, the target rate $\hat R_0$ is achievable for $U_0$ when $\log_2(1+P_0 g_0) \ge \hat R_0$; Otherwise, $U_0$ experiences an outage due to $\log_2(1+P_0 g_0) < \hat R_0$.

To enhance the transmission reliability, $U_1$ applies RS for each block of transmissions. Specifically, $U_1$ splits its signal $x_1$ into two parts $x_{1,1}$ and $x_{1,2}$ for transmissions \cite{Rate_Splitting_MAC,RS_NOMA_UL_fair}. At the end of each transmission block, the received signal at the BS can be expressed as
\begin{eqnarray}
    ~y =  \sqrt{P_0} h_0 x_0 + \sqrt{\alpha P_1} h_1 x_{11} + \sqrt{(1 \!-\! \alpha) P_1} h_1 x_{12} + n, \label{eq:y}
\end{eqnarray}
where $x_0$ is the transmit signal of $U_0$, $n$ is the Gaussian background noise modeled with a normalized power, and $\alpha$ is the power allocation factor satisfying $0 \le \alpha \le 1$. We assume that all the signals, $\{x_0, x_1, x_{11}, x_{12}\}$, are independently coded with Gaussian code book and each signal has a unit power in expectation. The signal model in \eqref{eq:y} is also called as rate splitting multiple access (RSMA) \cite{Rate_Splitting_MAC}. In this paper, we use the RS as in \eqref{eq:y} to improve the transmission robustness of the secondary user in the considered CR-NOMA system. According to \cite{Rate_Splitting_MAC}, the SIC decoding order $x_{11} \to x_0 \to x_{12}$ is applied at the BS receiver to achieve the benefit of RSMA.  Consequently, the received SNR/SINRs for decoding $x_{11}$, $x_0$, and $x_{12}$ can be expressed as
$
    \gamma_{11} = \frac{\alpha P_1 g_1}{ P_0 g_0 + (1-\alpha)P_1 g_1 + 1}     \label{eq:rate_rs1}
$,
$
    \gamma_0 =  \frac{P_0g_0}{(1-\alpha)P_1 g_1 + 1}
$,
and
$
    \gamma_{12} =  (1-\alpha)P_1 g_1
$,
respectively. For the primary and secondary users, the achievable rates are given by $R_0 = \log_2(1 + \gamma_0)$ and  $R_1 = R_{11} + R_{12}$ with $R_{11}= \log_2(1 + \gamma_{11})$ and $R_{12} = \log_2(1 + \gamma_{12})$, respectively.

\subsection{RS Strategy}

The RS strategy is designed to obtain the maximum allowed achievable rate $R_1$ for $U_1$, meanwhile keeping $U_0$ attaining the same outage performance as in OMA.
To begin with, we assume that $x_{11}$ has been detected correctly in the first SIC stage using the decoding order $x_{11} \to x_0 \to x_{12}$. Then, the remaining SIC is to detect $x_0$ and $x_{12}$, sequentially. When $\tau = 0$, the $U_0$'s transmission is always in outage and the detection of $x_{12}$ will fail in the remaining SIC with decoding order $x_0 \to x_{12}$. Nevertheless, when $\tau >0$ and $\gamma_{12} \le \tau$, it can be shown that $R_0 \ge \hat R_0$, so that the achievable rate becomes $R_1 = R_{11} + R_{12} $ for $U_1$. To maximize the achievable rate $R_1$, the RS strategy is designed to maximize $R_{12} = \log_2(1+ \gamma_{12})$ considering that the detection of $x_{12}$ is interference-free in SIC.

Depending on the different values of $P_1 g_1$ and $\tau$, the RS strategy is designed as follows:

1) Case I: $\tau > 0$ and $P_1 g_1 \le \tau$.
In order to maximize $R_1$, it is needed to allocate all the transmit power $P_1$ to transmit $x_{12}$, such that $R_{12} = \log_2(1+\gamma_{12})$ is maximized. Consequently, the RS strategy sets $\alpha = 0$ and $x_{12} = x_1$. Since   $x_{11}$ is not transmitted, the decoding order $x_{11} \to x_0 \to x_{12}$ degrades to $x_0 \to x_{12}$ (or equivalently $x_0 \to x_1$). Thus, the achievable rate for $U_1$ becomes
\begin{eqnarray}
R^{(\rm I)} =  \log_2\left(1 + P_1 g_1  \right).
\end{eqnarray}

2) Case II: $\tau > 0$ and $P_1 g_1 > \tau$. In this case, to maximize the achievable rate $R_1$,   $R_{12} = \log_2(1+\gamma_{12})$ needs to be maximized first subject to the constraint of $\log_2(1+\gamma_{12})~\le~\log_2(1+\tau)$, i.e., the transmission of $x_{12}$ cannot decrease the outage performance of $U_0$ for the transmission of $x_0$. Obviously, the allowed maximum $R_{12} = \log_2(1+\tau)$ is achieved by setting $\log_2(1+\gamma_{12}) = \log_2(1+\tau)$, which results in the optimal power allocation factor $\alpha = 1 - \tfrac{\tau}{P_1 g_1}$ for transmitting $x_{12}$ with the power $(1-\alpha) P_1$.
Meanwhile, the remaining transmit power, $\alpha P_1$, is allocated to transmit $x_{11}$, which yields the achievable rate $R_{11} = \log_2\left( 1 + \tfrac{P_1 g_1 - \tau}{P_0 g_0 + \tau + 1} \right) $. In this case, the target rates to transmit $x_{11}$ and $x_{12}$ are set by $\hat R_{11} = \hat R_1 - \hat R_{12}$ and $\hat R_{12} = \log_2(1 + \tau)$, respectively, and the achievable rate for the $U_1$'s transmission is given by
\begin{eqnarray}
R^{({\rm II})} =   \log_2\left( 1 + \tfrac{P_1 g_1 - \tau}{P_0 g_0 + \tau + 1} \right)   +  \log_2(1 + \tau).
\end{eqnarray}

To ensure that $U_0$ achieves the same outage performance as in OMA, it requires the successful detection of $x_{11}$ before detecting $x_0$. Nevertheless, $x_{11}$ cannot be detected correctly if $R^{(\rm II)} < \hat R_1 $. Therefore, in the proposed RS strategy, $U_1$ is allowed to transmit only if $R^{(\rm II)} \ge \hat R_1 $. Otherwise, $U_1$ keeps silent.

3) Case III: $\tau = 0$. In this case, the $U_0$'s transmission always encounters outage. To efficiently utilize transmit power, $U_1$ does not allocate any part of $P_1$ to transmit $x_{12}$ which cannot be detected correctly due to failure detection of $x_0$. To maximize the achievable rate $R_1$, the RS strategy sets $\alpha = 1$ and $x_{11} = x_1$. Consequently, the decoding order $x_{11} \to x_0 \to x_{12}$ degrades to $x_{11} \to x_0$ (or equivalently $x_1 \to x_0$), so that the achievable rate for $U_1$ becomes
\begin{eqnarray}
R^{(\rm III)} =  \log_2\left( 1 + \tfrac{P_1 g_1 }{P_0 g_0  + 1} \right).
\end{eqnarray}

Obviously, the difference between the proposed RS scheme and the QoS-SIC and NH-SIC schemes \cite{SGF_NOMA_QoS,Hybrid_SIC_NOMA_New} is that the QoS-SIC and NH-SIC schemes can attain the achievable rate in the form of $R^{(\rm I)}$ and $R^{(\rm III)}$, whereas only the  proposed RS scheme can attain the achievable rate $R^{(\rm II)}$ in Case II.

\section{Outage Analysis for the Proposed RS Scheme}

In the proposed RS scheme, the  primary user experiences the same outage performance as in OMA and the outage probability experienced by the secondary user can be expressed as follows:
\begin{eqnarray}
    P_{\rm out} = P_{\rm out}^{(\rm I)} + P_{\rm out}^{(\rm II)} + P_{\rm out}^{(\rm III)} ,
\end{eqnarray}
where
$
    P_{\rm out}^{(\rm I)} = \Pr\{ \tau > 0, P_1  g_1  \le \tau, R^{(\rm I)} < \hat R_1 \}
$,
$P_{\rm out}^{(\rm II)}~=~\Pr\{ \tau > 0, P_1  g_1  > \tau, R^{(\rm II)} < \hat R_1 \}
$,
and
$
    P_{\rm out}^{(\rm III)}~=~\Pr\{ \tau = 0, R^{(\rm III)} < \hat R_1 \}
$
denote the probabilities that $U_1$ is in outage corresponding to the RS scheme's three operation cases, respectively.

The closed-form expression for $P_{\rm out}^{(\rm II)} $ is provided in the following theorem.

\begin{theo}
    Corresponding to the RS scheme's operation of Case II, the probability that $U_1$ is in outage is given by
    \begin{eqnarray}
        P_{\rm out}^{(\rm II)} = e^{\frac{1}{P_1}} \mu\left( \frac{1}{P_1 \eta_0}\right) - e^{-\frac{\varepsilon_0 + \varepsilon_1 + \varepsilon_0 \varepsilon_1}{P_1}} \mu\left( -\frac{P_0}{P_1} \right)    \label{eq:poutII}
    \end{eqnarray}
    with $\eta_0 \triangleq \frac{\varepsilon_0}{P_0}$ and
    \begin{eqnarray}
        \mu(\nu) = \left\{ {\begin{array}{*{20}{c}}
            {\eta_0 \varepsilon_1 }, &{ {\rm if~~} \nu = 0 },\\
            \frac{e^{- \eta_0 (\nu+1) } - e^{- \eta_0 (\nu+1) (1+\varepsilon_1)  }}{ \nu + 1 } , &{\rm otherwise}.
            \end{array}} \right.
    \end{eqnarray}
\end{theo}
\begin{IEEEproof}
    See Appendix.
\end{IEEEproof}

Although the operation of the proposed RS scheme in Case II is different from those of the QoS-SIC and NH-SIC schemes \cite{SGF_NOMA_QoS,Hybrid_SIC_NOMA_New}, their operations in Cases I and III are the same. As a result, closed-form expressions for $P_{\rm out}^{(\rm I)} $ and $P_{\rm out}^{(\rm III)} $ can be retrieved from \cite{SGF_NOMA_QoS} and \cite{Hybrid_SIC_NOMA_New} as in the following theorem.

\begin{theo}
    Corresponding to the RS scheme's operations in Cases I and III, probabilities that $U_1$ is in outage are given by
    \begin{eqnarray}
        P_{\rm out}^{(\rm I)} = e^{-\eta_0} - e^{\frac{1}{P_1}} \mu\left( \frac{1}{P_1 \eta_0}  \right) - e^{-\eta_0(1 + \varepsilon_1) - \eta_1},
    \end{eqnarray}
    \begin{eqnarray}
        P_{\rm out}^{(\rm III)} = 1 - e^{-\eta_0} - \frac{ e^{-\eta_1} (1 - e^{-\eta_0(P_0\eta_1 + 1)}) }{P_0 \eta_1 + 1}  ,
    \end{eqnarray}
    respectively, where $\eta_1 \triangleq \frac{\varepsilon_1}{P_1}$.
\end{theo}

\begin{cor}
    The outage probability experienced by the secondary user in the proposed RS scheme is given by
    \begin{eqnarray}
        P_{\rm out} &\!\!\!=\!\!\!& 1 -  e^{-\eta_0(1 + \varepsilon_1) - \eta_1}
         - e^{-\frac{\varepsilon_0 + \varepsilon_1 + \varepsilon_0 \varepsilon_1}{P_1}} \mu\left( - \frac{P_0}{P_1}   \right) \nonumber \\
         &\!\!\! \!\!\!&  - \frac{ e^{-\eta_1} (1 - e^{-\eta_0(P_0\eta_1 + 1)}) }{P_0 \eta_1 + 1}. \label{eq:Pout}
    \end{eqnarray}
\end{cor}

\begin{IEEEproof}
By combing $P_{\rm out}^{(\rm I)} $, $P_{\rm out}^{(\rm II)} $, and $P_{\rm out}^{(\rm III)} $, \eqref{eq:Pout} can be derived.
\end{IEEEproof}

{{\emph{Remark 1:}}} In the high SNR region, i.e., $P_0 = P_1 \to \infty$, the probability terms can be approximated as $P_{\rm out}^{(\rm I)} \approx \eta_1 $, $P_{\rm out}^{(\rm II)}~\approx~ \frac{\varepsilon_0\varepsilon_1(1+\varepsilon_0)(1+\varepsilon_1)}{P_1^2}$, and $P_{\rm out}^{(\rm III)} \approx \frac{\varepsilon_0\varepsilon_1(1+ \varepsilon_0)}{P_1^2}$, which results in an approximation for the outage probability as $P_{\rm out} \approx \eta_1 $.

{{\emph{Remark 2:}}} Recalling the relationship between the upper and lower bounds on the channel gain $g_1$ for deriving $P_{\rm out,II}$ as in Appendix A, the constraint $\eta_0 < g_0 <  \eta_0 (1+\varepsilon_1) $ is required without imposing additional restriction on the values of $\varepsilon_0$ and $\varepsilon_1$, whereas the QoS-SIC scheme of \cite{SGF_NOMA_QoS} needs $\varepsilon_0 \varepsilon_1 <1$ to avoid the error floor in the outage probability. Consequently, the proposed scheme can achieve the diversity gain of 1 for all the feasible values of $\varepsilon_0$ and $\varepsilon_1$ as indicated by $P_{\rm out} \approx  \eta_1 $.

\section{Comparison Between The Proposed Scheme and Benchmark Schemes}

In this section, we compare the system performance of our proposed RS scheme with those of the QoS-SIC and NH-SIC schemes \cite{SGF_NOMA_QoS,Hybrid_SIC_NOMA_New}, respectively.

For the considered three schemes, the operations in Cases I and III result in the same probabilities $P_{\rm out}^{(\rm I)}$ and $P_{\rm out}^{(\rm III)}$ for $U_1$.
The differences among the three schemes only lie on the operation for the Case II. For Case II, the QoS-SIC and NH-SIC schemes attain the achievable rates $\tilde R^{(\rm II)} = \log_2\left(1 + \frac{P_1 g_1}{P_0 g_0 + 1}  \right)$ and $\bar R^{(\rm II)} = \max\left\{\log_2\left(1 + \frac{P_1 g_1}{P_0 g_0 + 1}  \right), \log_2(1+\tau)\right\}$, respectively  \cite{SGF_NOMA_QoS,Hybrid_SIC_NOMA_New}.

For the QoS-SIC and NH-SIC schemes, the probabilities that $U_1$ is in outage for Case II can be evaluated as
\begin{eqnarray}
    \bar P_{\rm out}^{(\rm II)} &\!\!\!=\!\!\!& \!\!\!\! \mathop{\mathcal{E}}\limits_{  g_0 >  \frac{\eta_0(1+\varepsilon_1)}{1 - \varepsilon_0 \varepsilon_1 }   } \!\! \left\{ \! \frac{P_0\varepsilon_0^{-1}g_0 \!-\! 1 }{P_1} \!<\! g_1 \!<\! \frac{\varepsilon_1(1 \!+\! P_0g_0)}{P_1} \! \right\} ~~~~
    \nonumber \\
    &\!\!\!=\!\!\!&\!\! \frac{ e^{\frac{1}{P_1} - \left(\frac{1}{P_1\eta_0} + 1  \right)\frac{\eta_0(1+\varepsilon_1)}{1-\varepsilon_0\varepsilon_1} } }{\frac{1}{P_1\eta_0} + 1}  \!-\! \frac{e^{-\eta_1 - (P_0\eta_1 + 1)\frac{\eta_0(1+\varepsilon_1)}{1-\varepsilon_0\varepsilon_1}     }   }{ P_0\eta_1 + 1  } ,~~~~~\label{eq:PoutII_QoS}
\end{eqnarray}
\begin{eqnarray}
    \tilde P_{\rm out}^{(\rm II)} &\!\!\!=\!\!\!&  \!\!\!\! \mathop{\mathcal{E}}\limits_{ \eta_0 < g_0 <  \eta_0(1+\varepsilon_1)  } \!\! \left\{ \! \frac{P_0\varepsilon_0^{-1}g_0 \!-\! 1 }{P_1} \!< \! g_1 \!<\! \frac{\varepsilon_1(1 \!+ \! P_0g_0)}{P_1} \! \right\} \nonumber \\
    &\!\!\!=\!\!\!& e^{\frac{1}{P_1}} \mu\left( \frac{1}{P_1 \eta_0}   \right)   -
    e^{-\eta_1} \mu\left( P_0\eta_1    \right), \label{eq:PoutII00}
\end{eqnarray}
respectively \cite{SGF_NOMA_QoS,Hybrid_SIC_NOMA_New}. Furthermore, $\bar P_{\rm out}^{(\rm II)} $ in \eqref{eq:PoutII_QoS} only exists for $\varepsilon_0 \varepsilon_1 < 1$. When $\varepsilon_0 \varepsilon_1 \ge 1$, $\bar P_{\rm out}^{(\rm II)} = \frac{P_0 P_1(\varepsilon_0\varepsilon_1-1)}{(P_0 \varepsilon_1 + P_1)(P_0 + \varepsilon_0 P_1)}  $.

By comparing \eqref{eq:PoutII00} with \eqref{ap:PoutII_2}, it can be shown that $ P_{\rm out}^{(\rm II)}$ is less than $\tilde P_{\rm out}^{(\rm II)}$ by
\begin{eqnarray}
 \Delta P_{\rm out}^{(\rm II)}  &\!\!\! =  \!\!\!& \!\!\! \mathop{\mathcal{E}}\limits_{ \eta_0 < g_0 <  \eta_0(1+\varepsilon_1)  }   \left\{  \frac{(1+\varepsilon_0)(1+\varepsilon_1) - (1+P_0g_0)}{P_1} \right.\nonumber \\
 &\!\!\! \!\!\!& ~~~~~~ ~~~~~~ ~~~~~~  \left. < g_1 < \frac{\varepsilon_1(1+P_0g_0)}{P_1}  \right\}   \nonumber \\
   &\!\!\! = \!\!\!&    e^{-\frac{\varepsilon_0 + \varepsilon_1 + \varepsilon_0 \varepsilon_1}{P_1}} \mu\left( -\frac{P_0}{P_1}   \right) - e^{-\eta_1} \mu\left( P_0\eta_1   \right).
\end{eqnarray}

For the considered schemes working for Case II, the conditional outage probability for $U_1$  can be expressed as $P_{\rm out,}^{\rm (II), con} = \frac{P_{\rm out}^{(\rm II)}}{\Pr\{\tau>0, P_1g_1 >\tau\}}$, $\tilde P_{\rm out}^{\rm (II), con} = \frac{\tilde P_{\rm out}^{(\rm II)}}{\Pr\{\tau>0, P_1g_1 >\tau\}}$, and $\bar P_{\rm out}^{\rm (II), con} = \frac{\bar P_{\rm out}^{(\rm II)}}{\Pr\{\tau>0, P_1g_1 >\tau\}}$, respectively, with $\Pr\{\tau>0, P_1g_1 >\tau\} = \frac{P_1 \eta_0 e^{-\eta_0}}{ 1 + P_1 \eta_0}$.

{{\emph{Remark 3:}}}
The proposed RS scheme can achieve a lower outage probability than that of the NH-SIC scheme due to $\Delta P_{\rm out}^{(\rm II)} \ge 0$. Furthermore, as $P_0 = P_1 \to \infty$, we have $ \Delta P_{\rm out}^{(\rm II)} \to 0$. As a result, the proposed scheme and NH-SIC scheme achieve the same outage probability of $U_1$ in the high SNR region.

{{\emph{Remark 4:}}}
An advantage of the  proposed RS scheme is that it always attains a greater achievable rate than the QoS-SIC and NH-SIC schemes, which can be exemplified  by the following examples.

{{\emph{Example 1:}}} $\log_2 (1 + \tau ) < \log_2 \left(1 + \frac{P_1 g_1}{P_0 g_0 + 1} \right)$. For this example, we set $g_0 = g_1 = 10$, $P_0 =1$, $P_1 = 10$, and $\hat R_0 = 2$ bits per channel use (BPCU), which results in $\varepsilon_0 = 3$ and $\tau = \frac{7}{3}$. For the QoS-SIC and NH-SIC schemes, the achievable rates of $U_1$ are $\tilde R^{(\rm II)} = \log_2\left(1 + \frac{100}{11} \right) \approx  3.335$ BPCU and $\bar R^{(\rm II)} = \max\{\log_2\left(1 + \frac{100}{11} \right) , \log_2(1 + \frac{7}{3})\} \approx 3.335$ BPCU, respectively. For the proposed scheme, the achievable rate is $R^{(\rm II)} = \log_2\left(1 + \frac{293}{40} \right) + \log_2(1 + \frac{7}{3}) \approx 4.794$ BPCU, which is much greater than those of the benchmark schemes.

\begin{figure}[tb]
    \begin{center}
    \subfigure[$U_0$ and $U_1$ have the same transmit SNR]{
    \includegraphics[width=3.1in]{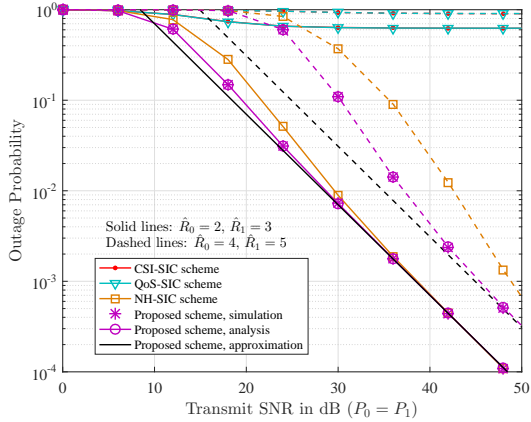}}
    \label{fig:subfig1a} \vspace{-0.15in}\\
    \subfigure[$U_0$ has a lower transmit SNR than $U_1$]{
    \includegraphics[width=3.1in]{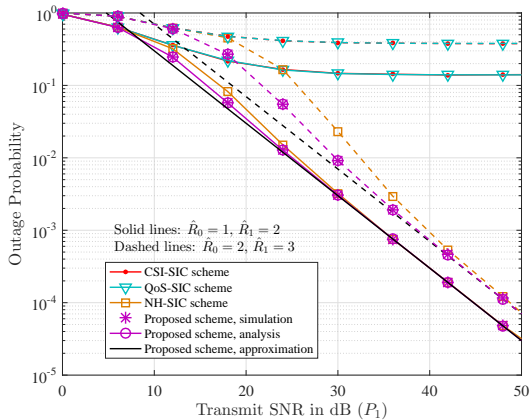}}
    \caption{Outage probability comparison under various transmit power settings.}
    \label{fig:subfig1b}
    \vspace{-0.2in}
    \end{center}
\end{figure}

{{\emph{Example 2:}}} $\log_2 (1 + \tau ) > \log_2 \left(1 + \frac{P_1 g_1}{P_0 g_0 + 1} \right)$. For this example, we set $g_0 = g_1 = 10$, $P_0 =10$, $P_1 = 20$, and $\varepsilon_0 = 3$. The interference threshold is determined as $\tau = \frac{97}{3}$. For the QoS-SIC and NH-SIC schemes, the achievable rate of $U_1$ are $\tilde R^{(\rm II)} = \log_2\left(1 + \frac{200}{201} \right) \approx  1.575$ BPCU and $\bar R^{(\rm II)} = \max\{ \log_2\left(1 + \frac{200}{201} \right) ,\log_2(1 + \frac{97}{3}) \}\approx 5.059$ BPCU, respectively. In contrast, the proposed scheme achieves $R^{(\rm II)} = \log_2\left(1 + \frac{503}{400} \right) + \log_2(1 + \frac{97}{3}) \approx 6.233$ BPCU, which is also greater than those of the benchmark schemes.

\section{Simulation Results}

In this section, we present the simulation results to clarify the outage performance achieved by the proposed RS scheme and verify the accuracy of the developed analytical expressions. For comparison purposes, the CSI-based SIC (CSI-SIC) scheme  is also considered \cite{NOMA_SIC}, in which the stronger user's signal is decoded before the weaker user's signal.

\begin{figure}[tb]
    \begin{center}
    \subfigure[Conditional outage probability comparison]{
    \includegraphics[width=3.1in]{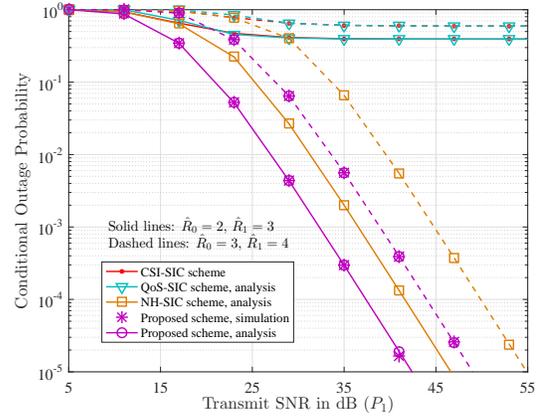}}
    \label{fig:subfig1a} \vspace{-0.15in}\\
    \subfigure[Impact of the target rate on the outage performance]{
    \includegraphics[width=3.1in]{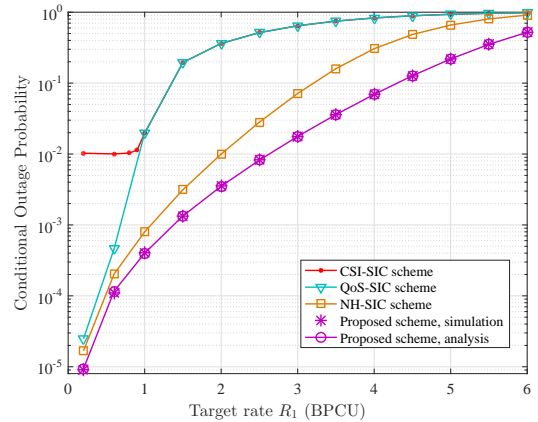}}
    \caption{Outage performance of $U_1$ under the Case II CSI condition.}
    \label{fig:subfig1b}
    \vspace{-0.25in}
    \end{center}
\end{figure}

In Fig. 1, the outage probability achieved by the proposed RS scheme is compared with the existing schemes for various transmit power settings. Particularly, we assume $P_0 = P_1$ in Fig. 1(a) and $P_0 = \frac{P_1}{10}$ in Fig. 1(b), respectively. The curves in Figs. 1(a) and 1(b) verify the accuracy of the analytical result based on Theorem 1 and the approximation expression for $P_{\rm out}$. Furthermore, we can see that the proposed scheme achieves the lowest outage performance among  the four schemes. Compared to the CSI-SIC and QoS-SIC schemes that cause the error floors on the outage probability, the proposed RS scheme achieves the outage probability monotonically decreasing with an increasing transmit SNR.
In addition, the proposed RS scheme achieves a lower outage probability than that of the NH-SIC scheme mostly in the middle-SNR region. In the high-SNR region, the proposed scheme and the NH-SIC scheme achieve almost the same outage probability due to $\Delta P_{\rm out}^{(\rm II)} \to 0$. Similarly to the NH-SIC scheme,  the proposed scheme avoids the error floor without imposing restriction on the values of $\varepsilon_0$ and $\varepsilon_1$.

\begin{figure}[tb]
    \begin{center}
    \includegraphics[width=3.1in]{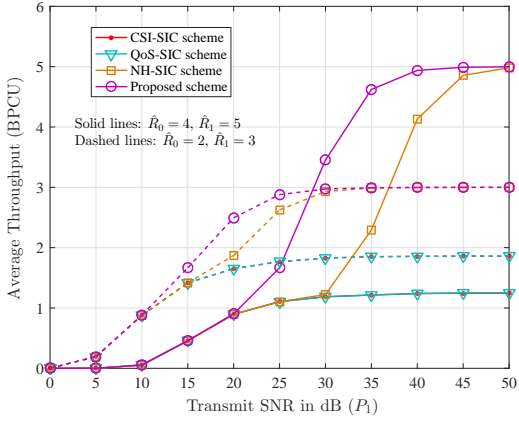}
    \caption{Average throughput comparison of the considered schemes ($P_0 = \frac{P_1}{10}$).}
    \label{fig:subfig3c}
    \end{center}
    \vspace{-0.3in}
\end{figure}

In Fig. 2, we investigate the conditional outage probability of $U_1$ achieved by the proposed RS scheme under the Case II  CSI condition. Based on the analytical results provided in Section III, we know that the outage performance differences among the existing schemes only rely on the operations for Case II. For the numerical results in Fig. 2(a), we set $P_0 = \frac{P_1}{10}$. In Fig. 2(a), the proposed RS scheme achieves the lowest conditional  outage probability. For the QoS-SIC scheme, the conditional outage probability achieved by it has an error floor in the middle- and high-SNR regions. However, the conditional outage probability values achieved by both the NH-SIC scheme and the proposed RS scheme monotonically decrease with an increasing  transmit SNR. The impact of the target rate on the probability that $U_1$ is in outage under the Case II CSI condition is investigated in Fig. 2(b), where we set a fixed target rate $\hat R_0 = 1$ BPCU in addition to $P_0 = 15$ dB and $P_1 = 20$ dB. The results in Fig. 2(b) show that the conditional outage probability achieved by all the considered schemes monotonically increases with an increasing target rate $\hat R_1$. Furthermore, the CSI-SIC scheme achieves the greatest conditional outage probability among the four schemes, while the conditional outage probability achieved by the QoS-SIC scheme is lower than that of the CSI-SIC scheme when $\varepsilon_0 \varepsilon_1 < 1$. Among the four schemes, the proposed RS scheme achieves the lowest conditional outage probability in the considered target rate region.

In Fig. 3, we evaluate the average throughput achieved by  $U_1$ to verify the superior performance of the proposed RS scheme. It is clearly illustrating that our RS scheme achieves the greatest average throughput among the existing ones. Compared to the CSI-SIC, QoS-SIC, and NH-SIC schemes, the average throughput achieved by the RS scheme is much greater in the middle and high SNR regions, which verifies that the proposed RS scheme exploits the transmit power more efficiently than the CSI-SIC, QoS-SIC and NH-SIC schemes.

\section{Conclusions}

In this paper, we have proposed an RS scheme to enhance the outage performance of the secondary user in the uplink CR-NOMA system without deteriorating the outage performance of the primary user compared to its counterpart in OMA. The proposed RS scheme can effectively allocate the transmit power to its split signal streams to improve the achievable rate in a CR-inspired way. The exact and approximated expressions for the outage probability of the secondary user have been derived and the numerical results have been provided, which clarify the superior outage performance of the proposed RS scheme over the existing SIC schemes.

\vspace{-0.05in}
\section*{Appendix: A proof of Theorem 1}

The probability term $P_{\rm out}^{(\rm II)}$ can be rewritten as
\begin{eqnarray}
    P_{\rm out}^{(\rm II)} &\!\!\!=\!\!\!& \Pr \left( g_0 > \eta_0, g_1 > \tfrac{P_0\varepsilon_0^{-1}g_0 - 1 }{P_1}, \right. \nonumber \\
    &\!\!\! \!\!\!&  \left. g_1 < \tfrac{(1+\varepsilon_0)(1+\varepsilon_1)-(1 + P_0g_0 ) }{P_1}  \right) \nonumber \\
    &\!\!\!=\!\!\!&
    \mathop{\mathcal{E}}\limits_{ \eta_0  <   g_0 < \eta_0(1+\varepsilon_1) + \frac{\varepsilon_1}{P_0}   }  \left\{ S \right\},   \label{ap:PoutII}
\end{eqnarray}
where $\eta_0 \triangleq  \frac{\varepsilon_0}{P_0}$, $\mathop{\mathcal{E}}\{\cdot \}$ stands for the expectation,  and $S$ is defined by
\begin{eqnarray}
S &\!\!\!  \triangleq \!\!\! & \Pr  \left(   \tfrac{P_0\varepsilon_0^{-1}g_0  -  1 }{P_1}  <  g_1  <  \tfrac{(1  +  \varepsilon_0)(1  +  \varepsilon_1)  -  (1 +  P_0g_0 ) }{P_1}   \right).~~      \label{ap:S}
\end{eqnarray}
Since $g_1$ is positive in practice, the inequality $ \frac{(1+\varepsilon_0)(1+\varepsilon_1)-(1 + P_0g_0 ) }{P_1}$  $> 0 $ holds true for \eqref{ap:PoutII}, which results in the constraint  $g_0 < \eta_0  (1+\varepsilon_1) + \frac{\varepsilon_1}{P_0}$ for the expectation in \eqref{ap:PoutII}. For the expression of $S$ in \eqref{ap:S}, the upper bound on $g_1$ should be greater than the lower bound on $g_1$, which results in the following hidden constraint
\begin{eqnarray}
    g_0 < \eta_0 (1 + \varepsilon_1).  \label{ap:hb_le_constraint}
\end{eqnarray}
Consequently, the expectation for $S$ in \eqref{ap:PoutII} should be taken over $ \eta_0  <   g_0 < \eta_0(1+\varepsilon_1)$ and $P_{\rm out}^{(\rm II)}$ can be evaluated as follows:
\begin{eqnarray}
    P_{\rm out}^{(\rm II)} &\!\!\!=\!\!\!&  \mathop{\mathcal{E}}\limits_{ \eta_0  <   g_0 < \eta_0(1+\varepsilon_1)  }  \left\{ S \right\}   \nonumber \\
    &\!\!\!=\!\!\!& \int\nolimits_{\eta_0 }^{\eta_0(1+\varepsilon_1)}  \int\nolimits_{\frac{P_0\varepsilon_0^{-1}y - 1 }{P_1}}^{\frac{(1+\varepsilon_0)(1+\varepsilon_1)-(1 + P_0y ) }{P_1} }   e^{-x} e^{-y}  dx dy. ~~~~~ \label{ap:PoutII_2}
\end{eqnarray}
After some integration manipulations, a closed-form expression for $P_{\rm out,II}$ can be readily obtained, which completes the proof.

\vspace{-0.08in}
\begin{balance}
\bibliography{IEEEabrv,IEEE_bib}
\end{balance}

\end{document}